\documentclass[fleqn,usenatbib]{mnras}

\usepackage{newtxtext,newtxmath}

\usepackage[T1]{fontenc}
\usepackage{ae,aecompl}

\usepackage{graphicx}	
\usepackage{amsmath}	
\usepackage{amssymb}	

\title[High-energy Cosmic Ray production in X-ray Binary Jets]{High-energy Cosmic Ray production in X-ray Binary Jets}

\author[A. J. Cooper et al.]{
A. J. Cooper,$^{1,2,4}$\thanks{E-mail: a.j.cooper@uva.nl}
D. Gaggero,$^{1,3}$
S. Markoff$^{1,2}$
S. Zhang$^{5,6}$
\\
$^{1}$GRAPPA, Gravitational and Astroparticle Physics Amsterdam, University of Amsterdam, Science Park 904, 1098 XH Amsterdam, the Netherlands\\
$^{2}$API, Anton Pannekoek Institute for Astronomy, University of Amsterdam, Science Park 904, 1098 XH Amsterdam, the Netherlands\\
$^{3}$Instituto de F\'isica Te\'orica UAM/CSIC, Calle Nicol\'as Cabrera 13-15, Cantoblanco E-28049 Madrid, Spain\\
$^{4}$ASTRON, Netherlands Institute for Radio Astronomy, Oude Hoogeveensedijk 4, 7991 PD, Dwingeloo, the Netherlands\\
$^{5}$Boston University, Institute for Astrophysical Research, 725 Commonwealth Avenue, Boston, MA 02215, USA\\
$^{6}$Bard College, Physics Program, 30 Campus Rd, Annandale-On-Hudson, NY 12504, USA
}

\date{Accepted XXX. Received YYY; in original form ZZZ}

\pubyear{2020}

\begin{document}
\label{firstpage}
\pagerange{\pageref{firstpage}--\pageref{lastpage}}

\maketitle

\begin{abstract}
As smaller analogs of Active Galactic Nuclei (AGN), X-ray Binaries (XRBs) are also capable of launching jets that accelerate particles to high energies.  In this work, we reexamine XRB jets as potential sources of high-energy cosmic rays (CRs) and explore whether they could provide a significant second Galactic component to the CR spectrum. In the most intriguing scenario, XRB-CRs could dominate the observed spectrum above the so-called ``knee'' feature at $\sim3\times10^{15}$ eV, offering an explanation for several key issues in this transition zone from Galactic to extragalactic CRs. We discuss how such a scenario could be probed in the near future via multi-messenger observations of XRB jets, as well as diffuse Galactic neutrino flux measurements.
\end{abstract}




\section{Introduction}

The origin of Cosmic Rays (CRs), high-energy particles from beyond the solar system, is a century-old puzzle \citep{Ginzburg1964,Ginzburg:1990sk,Blasi:2013rva}. We are yet to firmly identify classes of astrophysical sources able to accelerate hadronic cosmic particles up to extremely high energies; much larger than those accessible by terrestrial accelerators.
\par
Spectral features in the locally observed all-particle CR spectrum can shed light on this mystery. The observed spectrum follows a power law with an index of $p \approx -2.7$ over many decades of energy.
However, over years of observation, small deviations with respect to the power law have been identified as follows: \textit{the knee}, a softening of the spectra at $3 \times 10^{15}$ eV; the \textit{second knee}, a further softening at around $2 \times 10^{17}$ eV, and \textit{the ankle}, a hardening of the spectrum occurring at roughly $4 \times 10^{18}$ eV \citep{Blasi:2013rva}.
\par

It is commonly thought that Galactic sources are able to accelerate hadrons up to the knee, with Supernova Remnant (SNR) shocks as the prime candidates, yet many aspects of this picture are far from clear. CR paths are deflected in the Galactic magnetic field and therefore we cannot directly trace them back to their source. To this end, indirect ``smoking gun'' signals, including characteristic TeV $\gamma$-ray spectra from pion decay channels, can be observed to verify CR acceleration sites. Although observations of X-ray filaments \citep{2003ApJ...584..758V} and $\gamma$-ray spectra from old SNRs \citep{2013Sci...339..807A} suggest that protons are efficiently accelerated at these sources, it is not clear whether SNRs can universally attain the crucial PeV energies required to explain the softening at the knee (see e.g. the recent discussion in \citealt{2016GabiciGaggeroZandanel,2017MNRAS.472.2956A} and references therein). In the context of the SNR hypothesis, one of the key ideas to explain the knee and second knee features is rigidity-dependent diffusive shock acceleration (DSA), in which the maximum possible energy of a given CR species depends on its atomic number $Z$ such that $ E_{max}(Z) = Z * E_{max}(1)$. This has had success in explaining the second knee feature in terms of the cut-off of accelerated iron nuclei, as extensively discussed in the literature for many decades (see e.g. the early discussion in \citealt{Peters1961}). 
\par
\begin{figure}
  \centering
{\includegraphics[width=.5\textwidth]{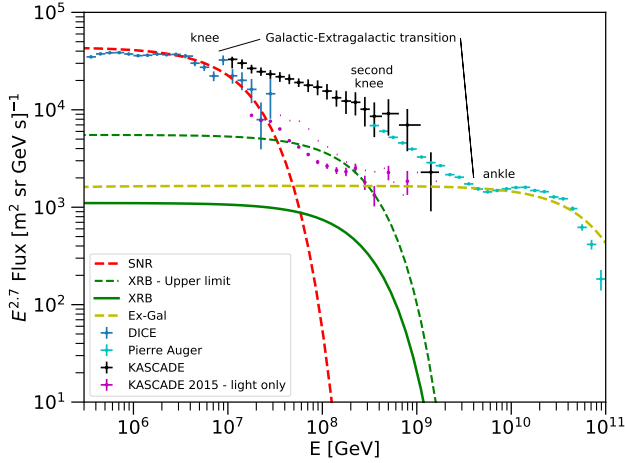}}
\caption{Schematic of a 3 source component all-particle cosmic ray spectrum. The components are SNR-CRs (red), XRB-CRs (this work; green) and a canonical extragalactic component (yellow).  The green line reflects an XRB-CR contribution with a total power of approximately $10^{38}$ erg/s, using the reasonable parameters in the middle column of Table \ref{tab:parameters}. The dashed green line represents the upper limit of the allowed XRB-CR power as discussed in Section 3, using the upper parameters in Table \ref{tab:parameters}. Such a contribution is dominant around $10^{16-17}$ eV, and could be probed via composition measurements. Here we have assumed all sources share the same powerlaw index of accelerated CRs. Allowing a slightly harder spectra for XRB-CRs mean they could explain the entire CR flux at the second knee without violating energetic constraints.}
\label{fig:CRSschematic}
\end{figure}

Besides the nature of the knee, we are still left with many open issues regarding the potential Galactic CR component. In particular, the origin of CRs between second knee and ankle, and the location of the transition from Galactic to extragalactic CRs remains unclear. 
One possibility is that the extragalactic component is dominant all the way down to $10^{17}$ eV, requiring the ankle feature to be a peculiar extragalactic propagation effect. However, alternative models also exist, mostly based on the assumption of energy-dependent leakage of high-energy cosmic rays from the Galaxy (see, for instance, \citealt{2015PhRvD..91h3009G}), which look to negate the need for an extragalactic CR component to dominate down to the second knee. Another option that has been put forward is the existence of a second Galactic component that ``fills the gap'' (see Fig. \ref{fig:CRSschematic}). Taking this additional component into account, as discussed e.g. in \cite{2005JPhG...31R..95H} and \cite{2013FrPhy...8..748G}, it is possible to provide a complete and consistent description of all the features from the knee to the ankle. Possible candidates for such high-energy Galactic components include strong Galactic winds \citep{Jokipii1987ApJ}, newborn pulsars \citep{Fang2013APS}, Galactic $\gamma$-ray bursts \citep{Levinson1993ApJ}, Wolf-Rayet star supernovae \citep{2016A&A...595A..33T} and many others.
\par
Recently, this idea of a second Galactic CR component has been bolstered by new composition measurements that favour a strong light composition at around $10^{17}$ eV. 
For instance, in \cite{2016Natur.531...70B}, a novel dataset based on 150 days of radio observations of CR-induced extensive air showers (EAS) made with the Low Frequency Array (LOFAR) implies a significant light-mass component in the $10^{17}-10^{17.5}$ eV range. The authors suggest this composition dip likely necessitates a primarily proton-dominated Galactic component which can reach approximately these energies. Measurements from different types of surface detectors including HiRes, Auger, Telescope Array, and KASCADE-Grande data (\citealt{2014arXiv1409.5083P,2019arXiv190801356H}; reviewed in e.g. \citealt{KAMPERT2012660}), are compatible with these recent findings, and clearly outline a decrease of the average mass of cosmic rays towards the ankle.

\par
In this paper we follow this line of inquiry and explore whether X-ray binary (XRB) jets, given the expected total overall power and maximum energy cut-off, could be viable candidates for the second Galactic source of CRs. Cosmic ray acceleration in XRB jets has been previously explored by \cite{2002A&A...390..751H} and later by \cite{2005MNRAS.360.1085F}, yet largely neglected since. In light of increasingly detailed multi-wavelength studies of many more XRB jets \citep{2016WATCHDOG,2016BLACKCAT}, which help constrain population statistics, as well as in anticipation of the next-generation of VHE $\gamma$-ray, neutrino and CR observatories, we revisit the possibility of CR production in XRB systems. We focus on the energy budget available for CR acceleration in all Galactic XRB jets and the maximum energy these XRB-CRs could attain, as these are the crucial inputs to determine a potential CR contribution. As mentioned, CRs are deflected in Galactic magnetic field and thus do not point back to their sources. To this end, we also investigate multi-messenger possibilities of verifying or falsifying XRB jets as a significant source of CRs. 

\section{X-ray Binary jets as cosmic ray accelerators}
Active Galactic Nuclei (AGN), also powered by accreting black holes, are a natural analogues to XRB jets and their similarities are starting to be quantifiable \citep{2003MNRAS.345.1057M,2004A&A...414..895F,2006Natur.444..730M,2012MNRAS.419..267P}. These systems have the theoretical capability to accelerate ultra high-energy CRs \citep{1984Hillas,2019MNRAS.482.4303M}, and we are beginning to see multi-messenger hints of extreme particle acceleration occurring either in the jets of AGN or at the termination shock sites \citep{2008APh....29..188P,2016Natur.531..476H,2018Sci...361..147I}. AGN jets are some of the prime candidates of the ultra high-energy extragalactic CRs and it is plausible that in the scaled down XRB jets we might expect similar CR production at lower energies, assuming similar physical processes occur across mass and luminosity scales. 
\par
Since their discovery as superluminal sources \citep{1994Natur.371...46M} XRB jets have been shown to accelerate leptons to very high energies in the jet-dominated hard state, where high-energy radiation is associated with extremely energetic electrons up to 100s TeV. The most characteristic examples are Cygnus X-1 \citep{2016A&A...596A..55Z}, Cygnus X-3 \citep{2009Natur.462..620T} and recently SS 433, which was resolved in the TeV range \citep{2018arXiv181001892H,2019arXiv191100013S}. Models of jet emission therefore require very high-energy electrons as sources of X-ray and  $\gamma$-ray emission (e.g \citealt{2005ApJ...635.1203M,2006A&A...447..263B,2014MNRAS.442.3243Z}). Shocks propagating in the jet likely accelerate charged particles to very high energies in a process known as diffusive shock acceleration (DSA) \citep{1977DoSSR.234.1306K,1978MNRAS.182..147B}, although other acceleration mechanisms such as magnetic reconnection could play a role \citep{2015MNRAS.450..183S}. Such particle acceleration may occur at any point along the jets, and the signature flat/inverted radio spectra suggests that continuous re-acceleration of radiating particles is required throughout the jet to combat adiabatic losses \citep{1979ApJ...232...34B,2014MNRAS.443..299M,2019MNRAS.482.2447P}.  Some authors have proposed specific zones offset from the black hole where this continuous acceleration initiates, such as near the base of the jet, or in a termination shock at jet-ISM working surface \citep{2002A&A...390..751H,2004MNRAS.355.1105F,2005ApJ...635.1203M,2009A&A...497..325B,2014MNRAS.439.1390R,2014MNRAS.442.3243Z}. 
\par
Although leptonic processes such as inverse-Compton scattering might be the dominant mechanism for such high-energy emission, hadronic particles may also significantly contribute. 
XRB jets are fed from accretion disks and stellar winds, presumably hadron-rich environments, yet the composition of XRB jets is still unclear. Observational evidence of Doppler shifted atomic line emission at relativistic velocities \citep{1979ApJ...233L..63M,Migliari1673,2013Natur.504..260D,2014A&A...571A..76D} suggests that at least some Galactic XRB jets have hadronic components, but whether this is ubiquitous is unknown. 
Protons/ions present in the jets will also undergo shock acceleration and in fact would attain much higher energies than electrons due their lower cooling efficiency compared to leptonic counterparts. Some authors suggest proton energies above $10^{15-16}$ eV are achievable in XRB jets, considering loss-limited acceleration due to radiative and adiabatic processes \citep{2008A&A...485..623R,2011arXiv1104.2071V,2015A&A...584A..95P}.
If efficient particle acceleration occurs in jets and a hadronic component is present, then a high-energy population of accelerated protons and ions is likely, making XRB jets promising candidate cosmic ray sources. 


\section{Population and CR power of Galactic XRBs}

The most important factors when considering generic CR sources are the total available CR power and the maximum attainable CR energy that the source can generate. The former relies on understanding the population and energetics of typical systems. While the latter cannot be directly determined at this time, recent improvements in the modeling of multi-wavelength data of XRBs is providing more realistic constraints on cooling rates, and thus potential CR energies.  We consider a model for the entire Galactic XRB population, and try to understand the potential total ensemble power. By considering the global energetics and estimating the proportion of power available for CR acceleration, we can estimate the CR flux and evaluate whether XRBs could reasonably account for a significant Galactic CR contribution. 
\par
XRBs come in four varieties, categorized by the compact object (either a black hole; BH, or neutron star; NS), and the companion star (low-mass; LM, or high-mass; HM). The outflows of each category of system depend strongly on the nature of their accretion. X-ray binaries with LM secondaries accrete via Roche lobe overflow, and thus undergo frequently recurring transient outbursts.
Thermal-viscous instabilities developing within the accretion disk give rise to outburst cycles \citep[e.g.,][]{2001NewAR..45..449L}. When observed in the X-ray waveband, a LM-XRB will evolve through a number of distinct accretion states defined by the source spectrum and luminosity \citep{2006RemillardMcClintock}. Comparatively, XRBs with HM secondaries tend to persistently accrete matter via strong stellar winds (with some notable exceptions; see e.g., \citealt{2016WATCHDOG}) and thus have somewhat more continuous outflows. The initial mass function (IMF; \citealt{1955ApJ...121..161S,2001MNRAS.322..231K}) states that low-mass stars are far more common than high-mass stars. Therefore BHs, which generally require more massive progenitors, are less common than NSs. Thus we expect many more NS-XRBs than BH-XRBs, as well as more systems with LM companions than HM companions. These distributions are encapsulated in population synthesis codes (e.g. \citealt{2008ApJS..174..223B,2011ApJS..192....3P,2015MNRAS.451.4086S}), which use the IMF, stellar evolution and binary interaction models to predict Galactic population statistics for each type of XRB.
\par
This being said, \cite{2005MNRAS.360.1085F} suggest that BH-LMXRB may actually dominate XRB-CR production despite the fact that they are less numerous in the Galaxy. The reason for this stems from the fact that their primarily due to their powerful radio jets implying large amounts of energy available. However, the number of BH-LMXRBs in the Milky Way is poorly constrained, with population synthesis predictions ranging from $10^{2}$ to $10^{5}$ (e.g. \citealt{2003ApJ...597.1036P,2006MNRAS.369.1152K,2006A&A...454..559Y,2008AIPC.1010..404S}). Unfortunately the large range of estimates is due to the uncertainties associated with modelling stellar evolution, particularly common envelope and SNe kick phases. 
\par
The most recent population synthesis results from \cite{2019arXiv190808775O} suggest $1.2 \times 10^{5}$ binaries involving a BH and main sequence star exist in the disk of the Milky Way. However, it is not immediately clear how many of these systems are actively transferring mass and could therefore be classed as XRBs. To understand this, we used the synthetic black hole catalog database provided by \cite{2019arXiv190808775O} \footnote{available at: https://bhc.syntheticuniverse.org/} to look at all binary systems containing a main sequence star and a black hole. For each of these systems, we looked up the binary separation, $a$, and approximated the radius of the main-sequence star from its mass. We estimate the proportion of the binaries which are actively transferring mass via Roche Lobe overflow by counting only those systems in which the radius of the main sequence star extends beyond the first Lagrangian point, $L_1$, of the system. We make use of the fitted formula of \cite{1964BAICz..15..165P} for the distance $b_1$ between $L_1$ and the centre of the primary:
\begin{equation}
    \dfrac{b_1}{a} = 0.5 - 0.227 \log(q)
\end{equation}
Where $q$ is the binary mass ratio. Given this criteria, we find 5531 XRBs in the model A datasets, and 5501 in the model B datasets, where the models differ slightly in the treatment of the common envelope phase. The vast majority of the XRBs are found in the Galactic disk. All the XRBs found in the datasets had main sequence stars of less than $10 M_{\odot}$ and so can be in general considered BH-LMXRBs.

\par
Recent observations suggest many more such systems may exist in the Galaxy than previously thought \citep{2016BLACKCAT,2016ApJ...825...10T,2018Natur.556...70H}. In particular, based on recent \textit{\textit{NuSTAR}} observations, \cite{2018Natur.556...70H} suggest that $300-1000$ BH-LMXRB might exist in just the central parsec of the Milky Way, and as such lower estimates from population synthesis simulations may be disfavoured. Given our population synthesis analysis, the density cusp in the Galactic centre and the uncertainties involved, we suggest $10^4$ is a reasonable upper limit for the Galactic XRB population, as reflected in Table \ref{tab:parameters}. While we take $10^{3}$ as a conservative lower estimate for the total number of Galactic BH-LMXRB in this work, we note that only $\sim 60$ BH-XRBs have been (observationally) confirmed to exist in the Galaxy \footnote{BlackCat BH-XRB catalog: http://www.astro.puc.cl/BlackCAT/}. Thus the true Galactic BH-LMXRB population remains a major source of uncertainty in our calculations. Furthermore, XRB outburst durations are typically on the order of months \citep{2016WATCHDOG}, yet CRs take Myrs to propagate through the Galaxy. Therefore predictions made based on current observations make the implicit assumption that the Galactic XRB population has not changed significantly in that time frame. 

\par
The fraction of an XRB jet's total power transferred to CR acceleration, the CR luminosity $L_\mathsf{CR}$, also involves many parameters lacking strict uncertainties. To estimate the realistic range of values of $L_\mathsf{CR}$, we use plausible ranges for each parameter. We adopt the method used in \cite{2005MNRAS.360.1085F}, using an outburst-oriented approach to incorporate advances in recent population studies, particularly the Watchdog database\footnote{available at: http://astro.physics.ualberta.ca/WATCHDOG/} \citep{2016WATCHDOG}. We simplify our calculations by only considering CR acceleration for BH-LMXRB systems in the hard, compact jet state\footnote{ Note that in \cite{2016WATCHDOG}, the hard, compact jet state is referred to as the ``Hard (Comptonized) State'' (HCS).}; as this is when we expect steady, particle accelerating outflows. By considering only the hard, compact jet state, we can obtain a conservative lower limit of $L_\mathsf{CR}$. Realistically, particle acceleration is also expected in other accretion states. In particular, the higher luminosity intermediate state, where transient jets and ejections are observed (e.g. \citealt{2012MNRAS.421..468M,2017MNRAS.469.3141T,2019ApJ...883..198R}), will likely contribute to $L_\mathsf{CR}$. However, given the comparatively short lifetime of the intermediate compared to hard accretion states in BH-LMXBs (i.e., $t_{IMS} \sim 1-10$ days, $t_{HS} \sim 20-50$ days; \citealt{2016WATCHDOG}), we do not believe the inclusion of the intermediate states will significantly alter our estimate of $L_\mathsf{CR}$.
\par
The following equation gives us an estimated total power of CRs, in units of the Eddington luminosity, from a set of N similar BH-XRB systems where $M_{BH} = 10 M_{\odot}$:

\begin{equation} \label{eq:1}
L_\mathsf{CR} \,=\, \dfrac{1}{2} \, \cdot \,\eta\, \cdot\,  \delta t \,\cdot\, A \,\cdot \, \bigg( \dfrac{L_\mathsf{X}}{L_\mathsf{edd}} \bigg)^\frac{1}{2} \,\cdot\, N
\end{equation}

Here, the factor of $\frac{1}{2}$ comes from the fact that we naively assume an equipartition between particles and magnetic fields, as well as sharing of the energy budget between leptons and hadrons. This gives us $\frac{1}{4}$ of the available power for hadronic acceleration, multiplied by two as there are two similar jets in each system. $\eta$ is the acceleration efficiency, i.e. how much of the jet power is transferred to high-energy particles via acceleration mechanisms, for which we take a canonical value of 0.1 as supported by simulations \citep{2014ApJ...783...91C}. As our model is based on outbursts, $\delta$t represents the hard state duty cycle of the systems. This factor is the average amount of time an BH-XRB spends in the hard state; the state in which we expect steady, compact jets which efficiently accelerate particles. 
\par
To estimate $\delta$t, we utilize the data collected in \cite{2016WATCHDOG}, in which the authors catalogued X-ray observations of all known BH-XRB over the last 19 years. In Table 15 of \cite{2016WATCHDOG}, we find detailed outburst statistics for 52 systems; 42 classed by the authors as transient, 10 as persistent. Most pertinent for our study, we find the number of days each system has spent in the hard, compact jet state, which is invaluable to constrain the hard state duty cycle. It is important to note that although almost 25\% of all BH-XRBs seem to be persistent accretors with high-mass companions, this is likely inflated due to observational bias due to their persistent and thus more reliably detected emission. As discussed high-mass companions are rarer and live shorter lives, and likely make up a minority of BH-XRB systems. 

\begin{figure}
  \centering
{\includegraphics[width=.5\textwidth]{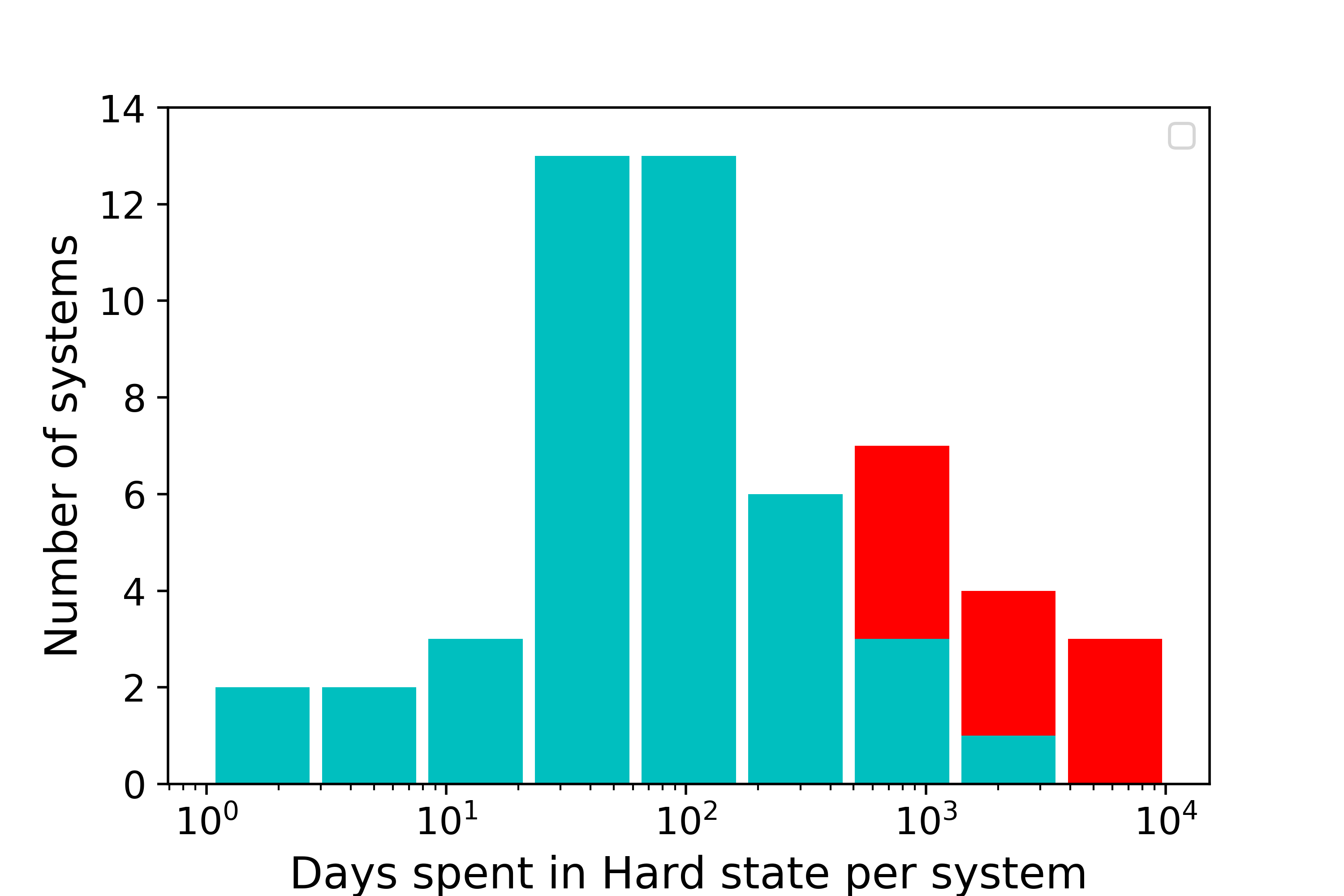}}
\caption{Histogram showing the number of days spent in the hard state for all known BH-XRBs over the entire 19 year period; in blue transient systems and in red persistent systems \citep{2016WATCHDOG}.}
\label{fig:XRBDutyCycle}
\end{figure}
\par

Of the transient systems, we find a mean and median number of days spent in the hard state of 183 and 66 days respectively. Taking the mean and median number of days divided by the total time in which the data was collected (19 years) as the duty cycle of these systems, we compute hard state duty cycles of 2.6\% and 1.0\% respectively. While this data set is the most complete to date with respect to XRB duty cycles, the mean duty cycle derived has to be considered as an upper limit, by virtue of the fact that only systems that have gone into outburst at least once are counted. Furthermore, the average outburst duration (months to years) is still somewhat comparable to the 19 year total on-time, which makes rigorous statistical statements difficult.

\par
Lastly, we use the $A$-parameter normalization prescription, as presented in \cite{2005MNRAS.360.1085F}, to evaluate the XRB jet power from X-ray observations. The A factor depends on both the type of XRB system and accretion state, and the values of A considered in Table \ref{tab:parameters} are chosen due to the discussion in \cite{2005MNRAS.360.1085F}. By combining (i) the relationship between jet power and radio luminosity motivated by models of steady, conical jets ($L_{\rm radio} \propto L_{J}^{\sim 1.4}$; \citealt{1979ApJ...232...34B,1995FalckeBiermann}) and (ii) the observed relation between X-ray and radio luminosity for accreting BH systems ($L_{\rm radio} \propto L_X^{0.7}$; \citealt{2000A&A...359..251C,2003MNRAS.345.1057M,2004A&A...414..895F}), \cite{2005MNRAS.360.1085F} was able to show that jet power ($L_J$) depends on the X-ray luminosity, according to:
\begin{equation}
L_J = A L_X^{0.5}
\end{equation}
Using an X-ray luminosity of jets varying between 1-5\% $L_{\rm edd}$ \citep{2003A&A...409..697M,2012MNRAS.421..468M,2014MNRAS.437.3265C,2019ApJ...883..198R}, the A-parameter normalization estimated by \cite{2005MNRAS.360.1085F} in this relation (see Table \ref{tab:parameters}), and a mean $\delta_t$ estimated from the WATCHDOG catalogue as discussed above, we are able to compute an $L_\mathsf{CR}$ estimate via Equation \ref{eq:1}.

\par
All together, we find a total XRB-CR power in the Milky Way of between approximately $10^{36}$ to $2 \times 10^{39}$ erg/s for the lower and upper bound parameters respectively, where each XRB provides an average CR power of $10^{33-36}$ erg/s depending on parameter choices. For the reasonable parameter values in the middle column of Table \ref{tab:parameters}, we find  $L_\mathsf{CR} \approx 10^{38}$ erg/s; approximately 1\% of the total estimated Galactic CR power. The actual XRB-CR power output is likely to be significantly higher as we neglect three important additional populations: persistently accreting BH-HMXRB systems, quiescent systems which are thought to behave much like jet-dominated hard state systems \citep{2013ApJ...773...59P}, as well as all NS-XRB systems, many of which have powerful jets \citep{2006MNRAS.372..417T,2006MNRAS.366...79M} which may accelerate CRs.

\begin{table}
\caption{Parameter limits for quantifying cosmic ray power of low-mass companion, black hole X-ray binary systems} 
\centering 
\begin{center}
\begin{tabular}{ |c|c|c|c| } 
\hline
Parameter & Upper & Middle & Lower \\
\hline
$A$ & 0.3 & 0.1 & $6 \times 10^{-3}$  \\ 
\hline
$\dfrac{L_X}{L_{edd}}$ & 0.05 & 0.03 & 0.01 \\
\hline
$N$ & $10^4$ & $3 \times 10^3$ & $10^3$\\ 
\hline
\end{tabular}
\end{center}
\label{tab:parameters}
\end{table}

\subsection{Constraints from Galactic centre observations}
Galactic CRs propagate from their sources interacting with interstellar gas to produce $\gamma$-rays and neutrinos. The observation of diffuse $\gamma$-ray emission in a region can therefore tell us about the density of both the ambient medium and high-energy CRs in that region. Furthermore, low-energy CRs interact with molecular clouds to produce X-rays. Observations of such clouds can be used to constrain the low-energy CR flux in the surrounding region. In the following, we look to the inner 200 parsecs of the Galaxy to constrain the power of Galactic XRB jets as CR sources. 
\par
\subsubsection{High-energy constraints on the CR power}
The recent \textit{NuSTAR} observation by \cite{2018Natur.556...70H} suggests the existence of a density cusp of BH-XRBs in the inner parsec of the Galactic centre. If this population of BH-XRBs is similar to the broader Galactic population in their potential to accelerate CRs, we expect to see $\gamma$-ray signatures of this in the region. Comparing the expected emission from CR-accelerating XRBs in the Galactic centre to the observed emission, we can constrain the CR power of these systems, and thus by extrapolation gain an additional constraint on the Galactic population as a whole. We use the very-high-energy $\gamma$-ray spectra observed by \cite{2016Natur.531..476H}, and assume that CRs accelerated in the jets of the \textit{NuSTAR} population of XRBs is responsible for all of the observed $\gamma$-rays. This is a very conservative constraint, as we assume all of the $\gamma$-ray emission is due to XRB-CR interactions with ambient protons. In reality, it is likely that many sources of CRs, including Sgr A* \citep{2016Natur.531..476H} and SNe in the region \citep{2017MNRAS.467.4622J}, and possibly other $\gamma$-ray production channels, contribute to the observed H.E.S.S. flux. In order to estimate the $\gamma$-ray emission from a population of cosmic particles injected by a cusp of XRBs located in the inner Galaxy, we perform both an analytical estimate and a numerical simulation. 
\par
For the analytical order-of-magnitude estimate, we follow the approach described in detail in \cite{2017MNRAS.467.4622J}. The authors consider the well-understood problem of a steady-state injection of hadrons at the GC from GeV all the way up to PeV energies with a single power-law energy spectrum, and their subsequent energy-dependent diffusive escape from a box with a Central Molecular Zone (CMZ) size $H$. They derive the following expression for the $\gamma$-ray luminosity associated to this hadronic population: 

\begin{eqnarray*}
L_{\gamma} ( > 200 {\rm GeV}) \,=\, 3.2 \times 10^{35} \left( \frac{H}{50\,{\rm pc}} \right)^2 \left( \frac{L_{\rm CR}}{1.6 \times 10^{39} \, {\rm erg/s}} \right) \times \\
\left(\frac{n}{100\, {\rm cm^{3}}} \right) \,\, {\rm erg/s}
\end{eqnarray*}

where $L_{CR}$ is the total power associated to the CR flux. Given the average gas density in the CMZ and a typical size of the region $H \simeq 100$ pc we get:

\begin{equation}
L_{\gamma} ( > 200 {\rm GeV}) \,=\, 1.2 \times 10^{36} \, {\rm erg/s} \, \left( \frac{L_{\rm CR}}{1.6 \times 10^{39} \, {\rm erg/s}} \right) 
\end{equation}

Given the $\gamma$-ray luminosity of the Galactic Ridge in $L_{\gamma}  \simeq 3.5 \times 10^{35} {\rm erg/s}$ as reported in \cite{2016Natur.531..476H}, it is straightforward to compute the maximum allowed power associated with the CR acceleration and injection in the ISM due to the population of the XRBs at the Galactic centre, that is still compatible with the H.E.S.S. measurement. The conservative upper limit on this quantity is $L_{\rm CR} \sim 10^{38} \, {\rm erg/s}$. For more details regarding the analytical estimate, we refer to Section 2 in \cite{2017MNRAS.467.4622J}. 
\par
In order to validate this estimate by means of a numerical simulation, we use the public codes {\tt DRAGON} \citep{2017JCAP...02..015E} and {\tt GammaSky}. Using these codes we are able to propagate CRs from any given source distribution and, adopting detailed models for the gas and interstellar radiation in the Galaxy, compute the $\gamma$-ray/neutrino flux associated to the CR population under consideration. We set up the {\tt DRAGON} code to inject CRs with a Gaussian source term centered on the Galactic centre with a $1$ pc width, consistent with the \cite{2018Natur.556...70H} population. We set a hard injection spectrum described by a single power law $Q = Q_0 (E/E_0)^{-\alpha}$ with $\alpha = 2.2$ and $E_{\rm min} = 1$ GeV, and let the particles propagate through the CMZ and diffuse out of the Galaxy. 

\par
\begin{figure}
  \centering
{\includegraphics[width=.5\textwidth]{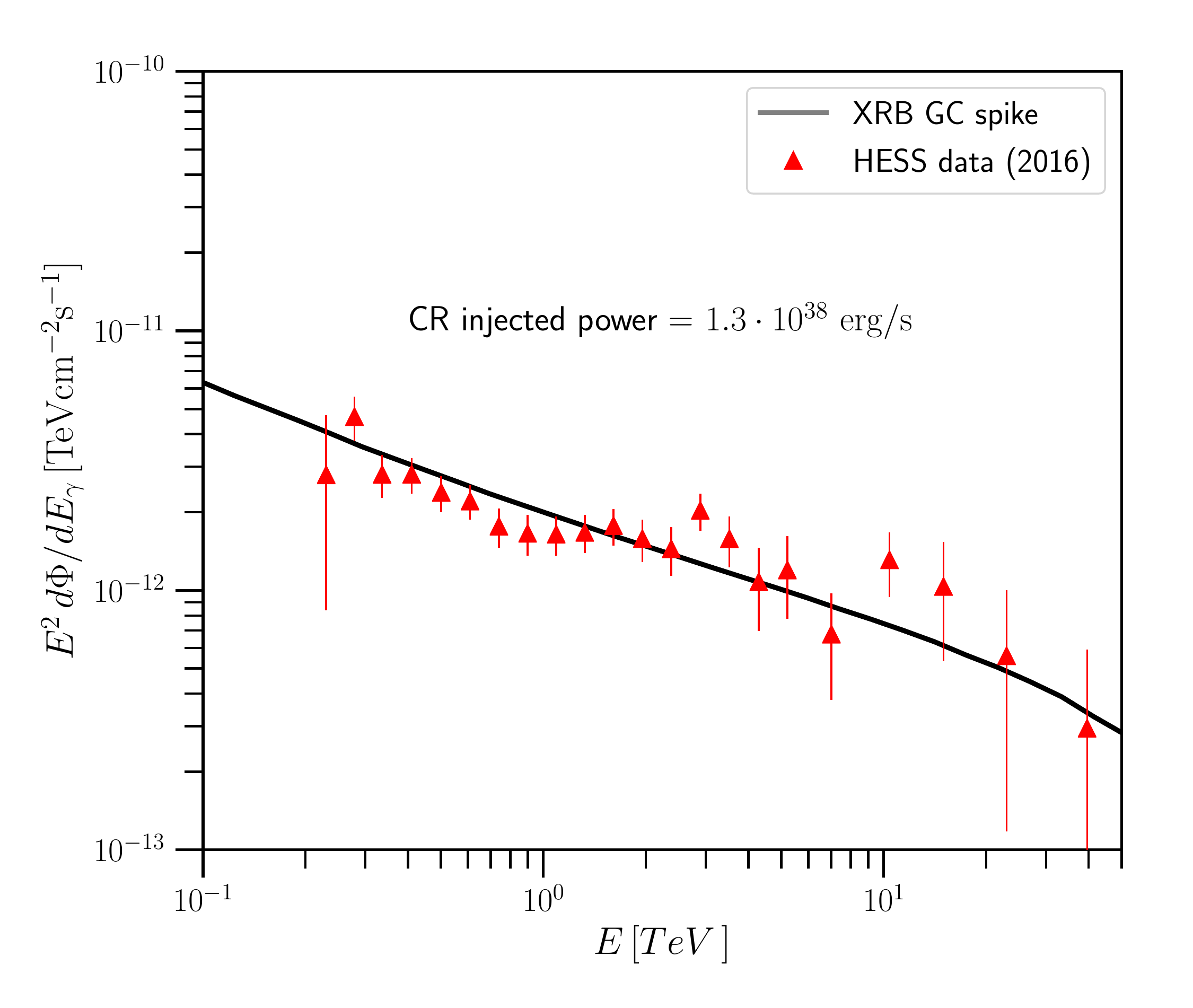}}
\caption{Gamma-ray spectral energy distribution associated to the population of CRs accelerated by XRBs located at the GC. We have assumed a CR injected power $L_{\rm CR}( > 1 {\rm GeV}) = 1.3 \times 10^{38} \, {\rm erg/s}$.  The $\gamma$-ray flux is integrated over the region of interest defined in \protect\cite{2016Natur.531..476H} (Figs 1 and 3); the H.E.S.S. data are shown as red triangles.}
\label{fig:HESSconstraint}
\end{figure}

After the equilibrium distribution of CRs is obtained, we compute the hadronic $\gamma$-ray flux from the Galactic Ridge region with the {\tt GammaSky} code, adopting the same model for the gas distribution in the CMZ as in \cite{2017PhRvL.119c1101G}. For a CR injected power $L_{\rm CR}( > 1 {\rm GeV}) \simeq 10^{38} \, {\rm erg/s}$, we obtain an average flux from the Galactic Ridge region ${\rm d}\Phi/{\rm d}E_{\gamma} = 2 \times 10^{-11}\,{\rm GeV^{-1} cm^{-2} s^{-1} sr^{-1}}$ at $1$ TeV (cfr. \citealt{2006NatureHESS}). Furthermore, we obtain an integrated flux of ${\rm d}\Phi/{\rm d}E_{\gamma} \simeq 2 \times 10^{-12}\,{\rm TeV^{-1} cm^{-2} s^{-1}}$ at 1 TeV from the inner annulus centered on Sgr A* as considered in \cite{2016Natur.531..476H}. Such $\gamma$-ray flux clearly saturates the $\gamma$-ray emission reported by the H.E.S.S. Collaboration (see Fig. \ref{fig:HESSconstraint}): thus we confirm the analytical order-of-magnitude estimate for the upper limit on the power injected in CRs at the GC from a XRB population.

\par
\cite{2018Natur.556...70H} suggest that between 300-1000 BH-XRBs exist in the Galactic centre. Although these systems currently seem to be mostly in quiescence, they could have been more active in the past. If we conservatively assume CR acceleration in the jets of these systems is responsible for all of the observed $\gamma$-ray flux in \cite{2016Natur.531..476H}, we can constrain the maximum CR power per system to be: $10^{34} \lesssim L_{\rm CR} \lesssim 3 \times 10^{35}$ erg/s. Extrapolating this to the wider Galactic population of $10^3 - 10^4$ systems, we find a total XRB-CR power of $10^{37} - 3 \times 10^{39}$ erg/s. This range of values falls within our estimates for the total XRB-CR power as found above; representing 0.1\% to 10\% of the total Galactic CR power. This is consistent with our total XRB-CR power derived earlier in this section. We stress that this is an upper limit based on the entire $\gamma$-ray flux as observed by \cite{2016Natur.531..476H} to originate from CRs accelerated in jets of the density cusp of XRBs in the Galactic centre, and is primarily used as a sanity check to ensure our assumptions do not violate observational limits.  

\par
\subsubsection{Low-energy constraints on the CR power}
An additional constraint on the CR power in the Galactic centre region comes from X-ray observations of the giant molecular clouds in the Central Molecular Zone (CMZ). Once again, this is an upper limit as we assume XRB-CRs from the \cite{2018Natur.556...70H} population are the only sources of CRs that contribute to the X-ray illumination. 
\par
GeV CR protons/ions bombarding giant molecular clouds produce X-ray emission through collisional ionization and bremssthalung. The Galactic centre molecular cloud Sgr B2's X-ray emission has been decaying over the last two decades, which is primarily due to X-ray echo of past activities of Sgr~A$^{\star}$ \citep{Inui2009, Terrier2010,Nobukawa2011}. However, in the recent years, as the X-ray echo component further faded away, Sgr~B2's X-ray emission seemed to enter a constant low flux stage, which is intepreted as GeV CR illumination. Observations of the Sgr B2 molecular cloud using \textit{NuSTAR} in 2013 have shown that, after more than ten years of flux decaying, the remaining X-ray emission from Sgr B2 is consistent with the GeV CR illumination scenario \citep{2015ApJ...815..132Z}. Assuming that all the Sgr B2 X-ray emission comes from CR illumination, \citet{2015ApJ...815..132Z} derived a CR proton spectral index of $\alpha=1.9^{+0.8}_{-0.7}$, and a required GeV proton power of $L_{\rm CR}=(0.4-2.3)\times10^{39}$ erg/s. We note that the GeV proton power derived from this method shall be taken as an upper limit, since the X-ray emission from Sgr B2 in 2013 can come partly from CR illumination, and partly from X-ray echoes. Future Sgr B2 X-ray observations will put a tighter constraints on the required CR proton power in the CMZ. This CR power upper limit from X-ray observation of Sgr B2 is consistent with the new estimate of XRB-CR power of the \cite{2018Natur.556...70H} population derived in this work.

\subsubsection{Consistency of constraints}
As we have both low-energy and high-energy constraints on the CR power in the region, we can check whether they are compatible by assuming our XRB-CR injection spectral index of $\alpha=2.2$ holds across the entire energy range. The total low-energy CR power upper limit of $(0.4-2.3) \times 10^{39}$ erg/s applies to CR energies between the $E_{\rm min} = 1-100$ MeV and $E_{\rm max} = 1$ GeV, i.e. the model parameters used by \citet{2015ApJ...815..132Z}. Using this, we extrapolate to find an allowed high-energy XRB-CR power due to the \cite{2018Natur.556...70H} population of $L_{\rm CR} ( > 1{\rm GeV}) \leq (1.3-39) \times 10^{38}$ erg/s, where the range of values reflects the allowed ranges of both the $E_{\rm min}$ parameter and low-energy CR power constraints described in Section 3.1.2. As the analytical and numerical analysis of the H.E.S.S. data suggests an upper limit of $L_{\rm CR} ( > 1{\rm GeV}) \sim 10^{38}$ erg/s, we find that the H.E.S.S. $\gamma$-ray measurements better constrain the CR power in the region, assuming $\alpha=2.2$. However, a softer CR injection spectra or alternative contributions to the observed $\gamma$-rays in region would mean the low-energy constraints are more stringent limits.

\section{Maximum Energy of XRB-CRs}

\begin{figure}
  \centering
{\includegraphics[width=.5\textwidth]{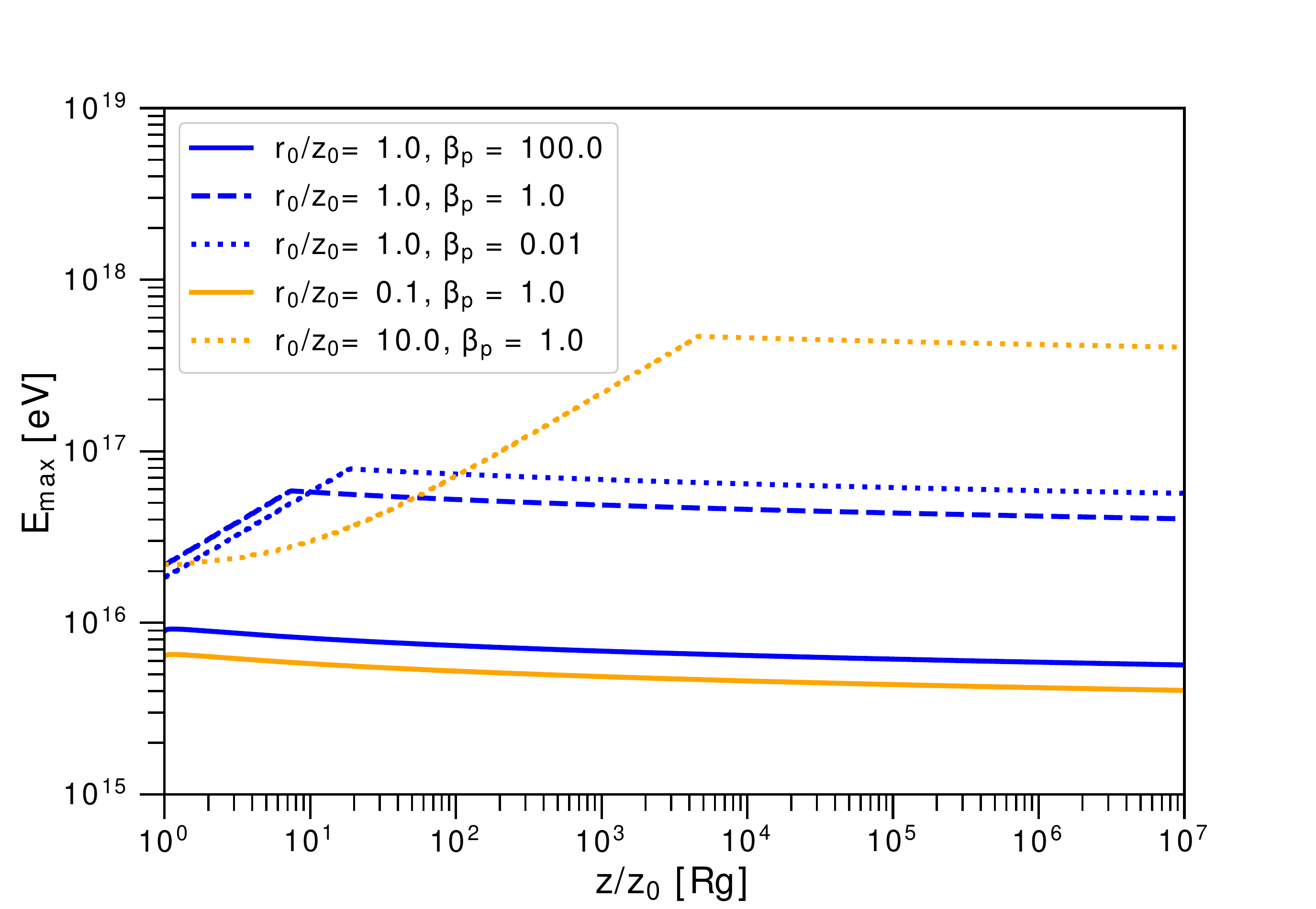}}
\caption{Maximum proton energy as a function of distance from the base of the jets for a quasi-isothermal jet model, where we use $\eta = 0.1$. We vary 2 important parameters: initial aspect ratio $\frac{r_0}{z_0}$ in orange and plasma $\beta_p = \dfrac{U_{e+p}}{U_{B}}$ in blue. The higher the initial aspect ratio, the wider the jet and thus the particles are confined more easily at high jet heights; the smaller $\beta_{p}$, the higher the magnetic field strength which results in smaller Larmor radii, aiding confinement, but producing larger synchrotron losses. In general, the maximum energy is limited by radiative losses at lower jet height due to strong magnetic fields, and is confinement-limited at large $z$.}
\label{fig:QuasiIsothermalMax}
\end{figure}
We have shown that BH-XRBs can viably contribute a significant fraction of the total Galactic CR power without violating constraints in the Galactic centre, so we now consider the maximum energy attainable by XRB-CRs. 
\par
The maximum energy of accelerated CRs in jets is limited either by energy losses (synchrotron, inverse-Compton and adiabatic losses are the primary channels) or by the Hillas criteria for confinement \citep{1984Hillas}. This is to say that accelerated CRs must stay confined within the accelerating medium in order to undergo re-acceleration, which we can quantify using the Larmor radius. The importance of the radiative losses can be quantified by comparing the timescales of the loss channels and the timescale of the acceleration mechanism. We compute the maximum energy as a function of jet height, $E_{\rm max}(z)$, such that it satisfies both of these constraints. Here we neglect proton-photon interactions. These interactions are expected to be sub-dominant at these extreme energies and magnetic field strengths even in photon-rich XRB systems \citep{2008A&A...485..623R,2015A&A...584A..95P,2019arXiv191100013S}. This is especially true for BH-LMXRB systems where the low-mass companions are expected to have a relatively modest photon field contributions.
\subsection{Jet Model}
We calculate the maximum energy of accelerated protons for the 3 different dynamical jet models (isothermal, adiabatic and quasi-isothermal \textit{agnjet} variant) outlined in \cite{2017A&A...601A..87C}. In the Appendix of this paper, we provide an overview of the different jet models and the parameters involved in computing the maximum CR energy. In Fig. \ref{fig:QuasiIsothermalMax}, we focus on the quasi-isothermal model as used in the \textit{agnjet} model, due to its ability to fit the flat jet spectrum we see in multi-wavelength XRB data \citep{2001A&A...372L..25M,2005ApJ...635.1203M}.
\par
In general, models of jets are based on the jet-disk symbiosis ansatz laid out in \cite{1995FalckeBiermann}. The jet is fed by the disk and the power of the jet at a height z is given by:
\begin{equation} \label{eq:5}
L_j(z) = \Gamma_j^2 \beta_j c \omega \pi z^2 \sin^2(\theta)
\end{equation}
where $\Gamma_j(z)$ is the lorentz factor of the bulk jet flow, $z$ is the height of the jet above the black hole ($z_0$ is the height of the jet base) and $\omega(z)$ is the enthalpy. For a jet with a co-moving particle number density, $n(z)$, the enthalpy can be written as:
\begin{equation}
\omega(z) = n m c^2 + U_j + P_j 
\end{equation}
Here, $U_j$ and $P_j$ are the energy density and the pressure of the jet respectively. We can approximate to:
\begin{equation} \label{eq:7}
    \omega(z) = n m_p c^2 + \Gamma_{\rm adi} U_j
\end{equation}
where we have assumed the jet can be treated as an ideal gas with adiabatic index, $\Gamma_{\rm adi}$, as in \cite{1995FalckeBiermann}. These equations are valid for all jet models considered in the Appendix. 
\par
To compute radiative losses and confinement of accelerated CRs in the jet, the most important parameters are the jet radius, $R(z)$, and the magnetic field strength, $B(z)$.  In all models, we define the magnetic field strength of the jet as:
\begin{equation}
    B(z) = \sqrt{\dfrac{8 \pi U_j(z)}{\beta_p + 1}}
\end{equation}
where:
\begin{equation}
   \beta_{p} = \dfrac{U_{e+p}}{U_B}
\end{equation}
$\beta_{p}$ is an important free parameter which sets how energy is distributed amongst particles and magnetic fields, and we show how different values of $\beta_{p}$ affect the maximum CR energy in Fig. \ref{fig:QuasiIsothermalMax}.
\par
The prescription of $U_j(z)$ depends on the choice of jet model as shown in the Appendix. From Equations \ref{eq:5} - \ref{eq:7}, we see that the value of $L_{j}(z_0)$, the power at the base of the jet, depends on $n(z_0)$ and $U_{j}(z_0)$, the number density and internal energy density at the base of the jet. In this analysis, we normalize $L_{j}(z_0)$ to the hard state jet power expected from the discussion in the previous section: $\sim 1-5 \%$ of the Eddington luminosity of  a $10 M_\odot$ black hole. This results in a jet base magnetic field strength, $B(z_0)$ of $(5-10) \times 10^{6}$ G, in line with other models \citep{2008A&A...485..623R,2015A&A...584A..95P}.
\par
To compute the radius of the jet, we follow \citep{2017A&A...601A..87C}. For the isothermal and adiabatic jet models, we use a simple conical jet model in which:
\begin{equation}
    r_{\rm cone}(z) = r_0 + (z - z_0) \sin(\theta)
\end{equation}
Here $\theta$ is the opening angle of the jet. This means that $r_0$ is an important free geometric parameter, which sets the initial radius of the jet. It directly influences the extent to which high-energy CRs can be confined, resulting in further acceleration. The quasi-isothermal \textit{agnjet} model used in Fig. \ref{fig:QuasiIsothermalMax} is not a conical model but instead considers self-collimation. This results in a slightly different jet radius profile:
\begin{equation} \label{eq:11}
    r_{\rm coll}(z) = r_0 + (z - z_0) \dfrac{\gamma_0 \beta_0}{\gamma_j \beta_j}
\end{equation}
This gives us a slightly narrower jet for larger values of $z$.
\par
When we calculate the maximum energy of accelerated CRs in the jet, we vary both $r_0$ and $\beta_{p}$. This helps us understand the parameter space available for a generic XRB population, and the different maximum CR energies attainable. As the magnetic field strength depends strongly upon the internal energy density of the jet, $U_j$, the maximum CR energy increases significantly for higher jet powers. This is to say that the most powerful XRB jets may be capable of producing higher energy CRs than outlined here. For more on the jet models please see the Appendix.
\par
\subsection{Calculating Maximum Energy}
The timescales of interest in computing the maximum energy are as follows:
\begin{equation}
t_{\rm acc}^{-1} = \dfrac{\eta e c B }{E}
\end{equation}
\begin{equation}
t_{\rm sync}^{-1} = \dfrac{4}{3} \Big(\dfrac{m_e}{m_p}\Big)^3  \, \dfrac{c \, \sigma_T \, U_B }{m_e c^2} \dfrac{E}{m_p c^2}
\end{equation} 
\begin{equation}
t_{\rm adi}^{-1} =  \dfrac{2}{3} \dfrac{\beta}{z}
\end{equation}
The maximum CR energy, as limited by radiation losses, is given by the condition:
\begin{equation} \label{eq:15}
t_{\rm acc}(E,z)^{-1} >  t_{\rm adi}(E,z)^{-1} + t_{\rm sync}(E,z)^{-1}
\end{equation}
The condition for confinement can be rewritten from Hillas' seminal paper \citep{1984Hillas} as: 
\begin{equation} \label{eq:16}
\textrm{E}_{\rm CR}(z) < \dfrac{B(z)}{\mu G} \times  \dfrac{R(z)}{pc} \times \dfrac{\beta}{0.5} \times 10^{15} \, \rm eV
\end{equation}
Equations \ref{eq:15} and \ref{eq:16} provide us with the constraints with which we compute the maximum jet power for all models. Specifically, we compute the maximum CR energy limited by each of these constraints, and take the minimum of these two values. In Fig. \ref{fig:QuasiIsothermalMax} we show our results for how the maximum possible energy varies as a function of the jet height for the quasi-isothermal jet model (that provides the best description of flat spectra jets; \citealt{2017A&A...601A..87C}), for different values of the initial aspect ratio, $\dfrac{r_0}{z_0}$, and $\beta_{p}$ of the jet. 
\par
Radiative losses dominate near the black hole, as the high magnetic field strength close to the base of the jet results in large synchrotron losses. Most models then show a flattening when a lack of confinement of the particles limits the maximum energy of XRB-CR higher up in the jet. One can assume that in the confinement-limited region, accelerated CRs which exceed the critical energy at which the particle stays confined escape the jet to propagate through the ISM.
\par
We find that the maximum attainable CR energy depends strongly on geometry, jet model and acceleration region; but in general protons can reach energies of $10^{16-17}$ eV if accelerated with a canonical efficiency of $\eta = 0.1$ \citep{2014ApJ...783...91C}. We note that varying $\eta$ scales the maximum energy linearly in the radiative loss dominated regime at small $z$. These calculations assume protons (i.e. $Z = 1$), but if more massive CR ions are present in the jet they would attain greater energies as the maximum CR energy scales with rigidity. Lastly, we note that in specific geometries and acceleration regions CR energies higher than $10^{17}$ eV could be reached, but this might only be plausible in atypical systems such as very powerful, wide or highly magnetized jets.

\section{Multi-messenger tests of the XRB-CR scenario}
Any source class contribution to the CR spectrum can only be directly probed by CR observatories if those sources dominate the spectrum at specific energies. Although XRB-CRs might dominate the parts of the CR spectrum, this is highly dependent on the total CR power, maximum energy of individual CRs and the acceleration powerlaw index. Given our results, we suggest that in the most optimistic case XRB-CRs might dominate (or contribute significantly to) the spectrum close to $10^{17}$ eV, near the second knee, where a light-mass component has been detected \citep{2014arXiv1409.5083P,2016Natur.531...70B,2019arXiv190801356H,2019ICRC...36..482Y}.
\par
The latest results from CR instruments seem to only strengthen the evidence for a light-mass component above $10^{17}$ eV \citep{2019ICRC...36..306K,2019ICRC...36..482Y}, and upgrades of such instruments  (e.g. \citealt{2019ICRC...36..363M,2016arXiv160403637T}) will be crucial to understand the composition of the transitional energy region between Galactic and extragalactic CRs. While this observed lighter mass component could be interpreted as the start of the extragalactic component, this would require the ankle to be a propagation effect. Thus any Galactic CR accelerator able to reach these energies is of great interest. However, the allowed range of XRB-CR power found in this work means that the contribution could be subdominant at all energy ranges, and thus any confirmation of CR acceleration in XRB jets might only be found via indirect measurements of $\gamma$-rays or neutrinos. In Fig. \ref{fig:CRSschematic}, we show a schematic of the all-particle CR spectrum, with a range of allowed contributions from the XRB-CR component calculated in this work. In particular, we show (green dashed line) the maximum allowed contribution, which is calculated by taking the upper parameters in Table \ref{tab:parameters}. Such a contribution would make up a significant fraction of the CRs in the energy range between second knee and ankle, where the role of a second Galactic component is currently under debate.

\subsection{$\gamma$-rays}
Several XRB jets are now known to emit $\gamma$-rays \citep{2009Natur.462..620T,2041-8205-807-1-L8,2016A&A...596A..55Z,2018arXiv181001892H}, although some observations have reported non-detections \citep{2013ApJ...775...98B,2017MNRAS.471.1688A,2018A&A...612A..14M}. Given the transient nature of some of these sources (and especially the complex environment of the SS 433 system), we do not necessarily expect CR acceleration and subsequent $\gamma$-ray emission continuously from XRB systems. Furthermore, although the observation of such high-energy radiation is a clear signature of particle acceleration, it is not trivial to pin down the origin of observed $\gamma$-rays which could be leptonic, hadronic or a combination. The upcoming, next generation Cherenkov Telescope Array (CTA; \citealt{2019scta.book.....C}) will have an order of magnitude better sensitivity compared to current facilities, and up to 4-5 orders of magnitude better sensitivity than \textit{Fermi} in the 100 GeV range for fast transients. CTA will thus likely be able to detect and identify the Galactic PeVatron sources in the near future (\textcolor{blue}{Kantzas et al. in prep}). 
\par
As a consistency check, we again used the {\tt DRAGON} code to look at the expected diffuse Galactic $\gamma$-ray emission due to the XRB population. However, as this population is sub-dominant to the (SNR) low-energy CR sources below $10^{16}$ eV, it is impossible to distinguish the sources in currently observable $\gamma$-ray wavelengths. Therefore, we suggest point source $\gamma$-ray observations of the most powerful XRB jets will pave the way for identifying CRs from XRBs through traditional electromagnetic observations. 

\subsection{Neutrinos}
Neutrinos are also produced through CR interactions with protons or photons, and XRB jets have long been predicted as a sources of neutrinos \citep{2001PhRvL..87q1101L,2002ApJ...575..378D}. As neutrino astronomy is still in its infancy, observations of Galactic-origin neutrinos thus far have been compatible with background \citep{1475-7516-2017-04-019}. However, the current limits from a joint analysis of ANTARES and IceCube data \citep{2018ApJ...868L..20A} are now getting close to the most optimistic predictions regarding the expected Galactic neutrino flux. Therefore, the clear detection of a component associated to the Galactic plane may be round the corner (see for instance a recent $2\sigma$ hint reported in \citealt{2019arXiv190706714A}), and diffuse Galactic searches could provide indeed a novel approach towards identifying a second source of Galactic CRs. Neutrino observations probe higher energies than $\gamma$-ray facilities and therefore high-energy breaks in the diffuse Galactic neutrino spectra \citep{2017ApJ...849...67A,2018ApJ...868L..20A} could be interpreted as separate contributions from different CR source classes.
\par
\begin{figure}
  \centering
{\includegraphics[width=.48\textwidth]{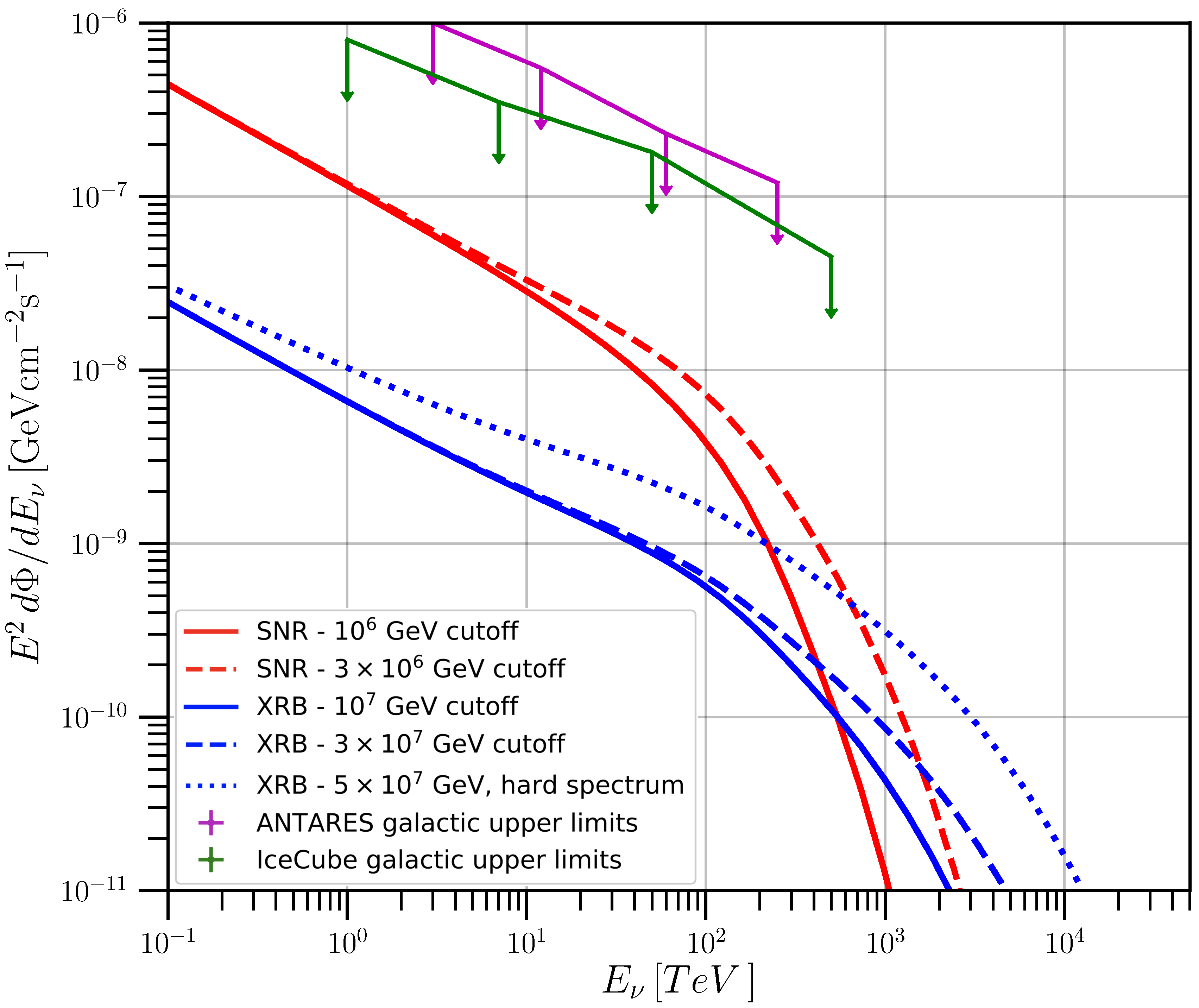}}
\caption{The predicted diffuse Galactic neutrino spectrum from SNR-CRs and XRB-CRs with joint upper limits from ANTARES and IceCube \citep{2017ApJ...849...67A} using the {\tt DRAGON} code. Specifically, we note that breaks in the spectra are predicted in the total spectrum at model-dependent sensitivities even with very conservative maximum energy cut-offs.}
\label{fig:NeutrinoSpectra}
\end{figure}

Using the {\tt DRAGON} code, we compute the expected diffuse neutrino emission due to CRs propagating from two different components: the dominant low-energy (SNR) component and a higher energy (XRB) component. We assume a low-energy component that saturates the observed CR spectrum below the knee, as expected by the dominant Galactic CR source. The subdominant, higher maximum energy component has approximately 10\% of the power of the low-energy source, in agreement with most optimistic findings of total XRB-CR power in Section 3. For the source distributions, we have assumed a Lorimer pulsar distribution \citep{2006MNRAS.372..777L} for both source populations, as this is a good approximate tracer of compact objects and therefore of SNRs and XRBs. In Fig. \ref{fig:NeutrinoSpectra}, we see the resulting diffuse neutrino spectra due to these two components, where we assume the low-energy and higher energy components are due to SNR and XRB sources respectively. We plot a different maximum energy cut-offs, as the maximum CR energy for each source is not well-known. Unlike cosmic rays, neutrinos trace back to their sources and thus confirming a break/hardening in the spectrum towards the Milky Way could verify the Galactic origin of high-energy CRs. The current upper limits on the Galactic contribution to the astrophysical neutrino flux by IceCube and ANTARES are also shown in the figure.
\par
As the next-generation of neutrino observatories come online (Icecube-Gen2, \citealt{2014arXiv1412.5106I}; KM3-NET (2.0), \citealt{2016JPhG...43h4001A}), we can probe PeV energy ranges in order to verify whether there are two clear populations of high-energy CR sources within our Galaxy. As diffuse neutrino limits are already encroaching on best models of neutrino emission from Galactic CRs, the ten-fold detector volume increase specified for IceCube-Gen2 will probe our predictions of a break in the Galactic neutrino spectrum due to a high-energy Galactic component. Furthermore, KM3-NET upgrades over the next decade will increase the angular resolution of detections to $<0.1^{\circ}$ at PeV energies. Coupled with greater sensitivities, point source neutrino astronomy will soon be at the forefront of identifying CR sources within our Galaxy. Once these upgrades are realised, XRB systems such as Cygnus X-1 will be key targets for neutrino observatories to test whether XRB jets are important CR accelerators. 

\section{Conclusion}
We have suggested that XRB jets could accelerate protons to high energies, similar to their larger counterparts in AGN. Within the uncertainties allowed by current population models, jet composition and Galactic centre observational constraints, a total XRB-CR power of between $10^{36-39}$ erg/s is possible. The most likely allowed value of around $10^{38}$ erg/s means XRB-CRs could contribute a few percent of the dominant SNR-CR component, representing a non-negligible contribution to the observed CR spectrum. The maximum energy of XRB-CR is relatively high compared to other Galactic sources of CRs, with models suggesting protons could be accelerated to $10^{16-17}$ eV in some systems. Together these two results indicate that XRB-CRs could even dominate the total CR spectrum in part of the transition region between SNR-CR and extragalactic CR components, above the knee and below the ankle, in broad agreement with recent mass composition results. Lastly, we suggest multi-messenger possibilities to confirm XRB-CR (or generic second Galactic components) through diffuse neutrino and $\gamma$-ray measurements of our Galaxy, as well as point source observations. 

\section*{Acknowledgements}
We would like to thank R. Bartels, S. Gabici, D. Kantzas and B. Tetarenko for helpful discussions. We would also like to thank the anonymous referee for their thorough and insightful comments which improved this work. 
\par
AC is partially
supported by the Netherlands Research School for Astronomy (NOVA). The work of DG has received financial support through the Postdoctoral Junior Leader Fellowship Programme from la Caixa Banking Foundation (grant n.~LCF/BQ/LI18/11630014). DG was also supported by the Spanish Agencia Estatal de Investigaci\'{o}n through the grants PGC2018-095161-B-I00, IFT Centro de Excelencia Severo Ochoa SEV-2016-0597, and Red Consolider MultiDark FPA2017-90566-REDC.  S.M. is supported by an NWO (Netherlands Organisation for Scientific Research) VICI award, grant Nr. 639.043.513.

\section*{Appendix}
In Section 4, we looked at dynamical jet models in order to estimate the maximum attainable CR energy as a function of jet height. We rely heavily on \cite{2017A&A...601A..87C}, in which the equations governing the different jet models are derived. Here we give a quick overview of each jet model, and show how the maximum CR energy depends on the jet model. As mentioned in Section 4 we believe that for XRB jets, the quasi-isothermal jet model is the most appropriate. 

\par
From this starting point, assumptions about the physics of the jet lead to different models. The most important difference is that in the adiabatic jet model, adiabatic losses are not compensated for. In all other models, losses are compensated for by e.g. continuous reacceleration of particles. In the isothermal model adiabatic losses are fully compensated for; whereas in the quasi-isothermal models only longitudinal ($z$-direction) losses are compensated for. These assumptions lead to different internal energy profiles, and different Euler equations from which the Lorentz profile of the jet is derived. In each case, we briefly explain the assumptions and list the Euler equation for the model. For a more thorough explanation, we suggest the reader refers to \cite{2017A&A...601A..87C}. 

\subsection*{Adiabatic}
In the adiabatic jet model, the jet conserves energy such that it obeys the relativistic Bernoulli equation: $\gamma_j \frac{\omega}{n} =$ const. This means that $T_j \propto n^{\Gamma_{\rm adi} - 1}$. The internal energy density profile is:
\begin{equation}
        U_j(z) = \zeta n_0 m_p c^2 \bigg(\dfrac{\gamma_j \beta_j}{\gamma_0 \beta_0}\bigg)^{-\Gamma_{\rm adi}} \bigg( \dfrac{z}{z_0}\bigg)^{-2 \Gamma_{\rm adi}}
\end{equation}
The Euler equation then is:
\begin{equation}
    \Bigg( \gamma_j \beta_j \dfrac{\Gamma_{\rm adi} + \xi }{\Gamma_{\rm adi} - 1} - \Gamma_{\rm adi} \gamma_j \beta_j - \dfrac{\Gamma_{\rm adi}}{\gamma_j \beta_j} \Bigg) \dfrac{\partial \gamma_j \beta_j }{\partial z} = \dfrac{2}{z}
\end{equation}
\begin{equation}
    \xi = \dfrac{1}{\zeta}  \Bigg( \gamma_j \beta_j \sqrt{\dfrac{1 + 2 \zeta \Gamma_{\rm adi} - \zeta \Gamma_{\rm adi}^2}{\zeta \Gamma_{\rm adi} (\Gamma_{\rm adi} - 1)}} \; \Bigg)^{\Gamma_{\rm adi}-1} \bigg( \dfrac{z}{z_0}\bigg)^{2 (\Gamma_{\rm adi} -1)}
\end{equation}

In terms of CR acceleration, adiabatic jets generally attain lower CR energies compared to other jet models, espcially at large $z$. This is because the internal energy density and thus magnetic field strength decreases rapidly as $z$ increases, as no reacceleration occurs. We stress that the adiabatic jet model cannot fit the flat spectra we see in XRB jets, and is presented primarily for comparison.

\begin{figure}
  \centering
{\includegraphics[width=.5\textwidth]{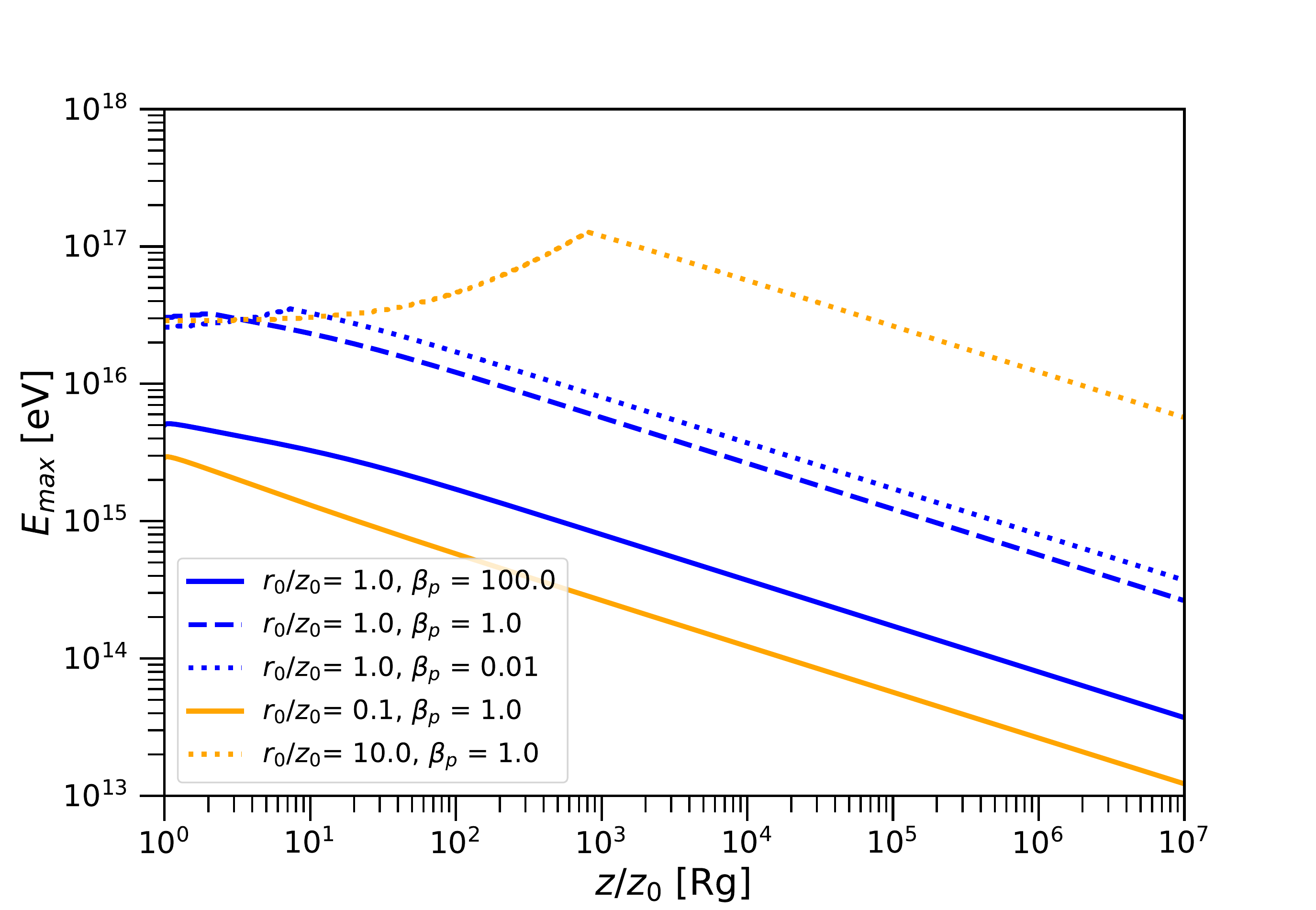}}
\caption{Maximum CR energy as a function of jet height $z$ for the adiabatic Jet model.}
\label{fig:AdiabaticJetMax}
\end{figure}

\subsection*{Isothermal}
In the isothermal jet model, all adiabatic losses are recompensated for and thus $T_j$ is constant. This means that $U_p \propto n$, and energy is not conserved.  
\begin{equation}
    U_j(z) = \zeta n_0 m_p c^2 \bigg(\dfrac{\gamma_j \beta_j}{\gamma_0 \beta_0}\bigg)^{-\Gamma_{\rm adi}} \bigg( \dfrac{z}{z_0}\bigg)^{-2}
\end{equation}
The Euler equation is:
\begin{equation}
    \Bigg( \gamma_j \beta_j \dfrac{\Gamma_{\rm adi} + 1 }{\Gamma_{\rm adi} - 1} - \Gamma_{\rm adi} \gamma_j \beta_j - \dfrac{\Gamma_{\rm adi}}{\gamma_j \beta_j} \Bigg) \dfrac{\partial \gamma_j \beta_j }{\partial z} = \dfrac{2}{z}
\end{equation}
\begin{figure}
  \centering
{\includegraphics[width=.5\textwidth]{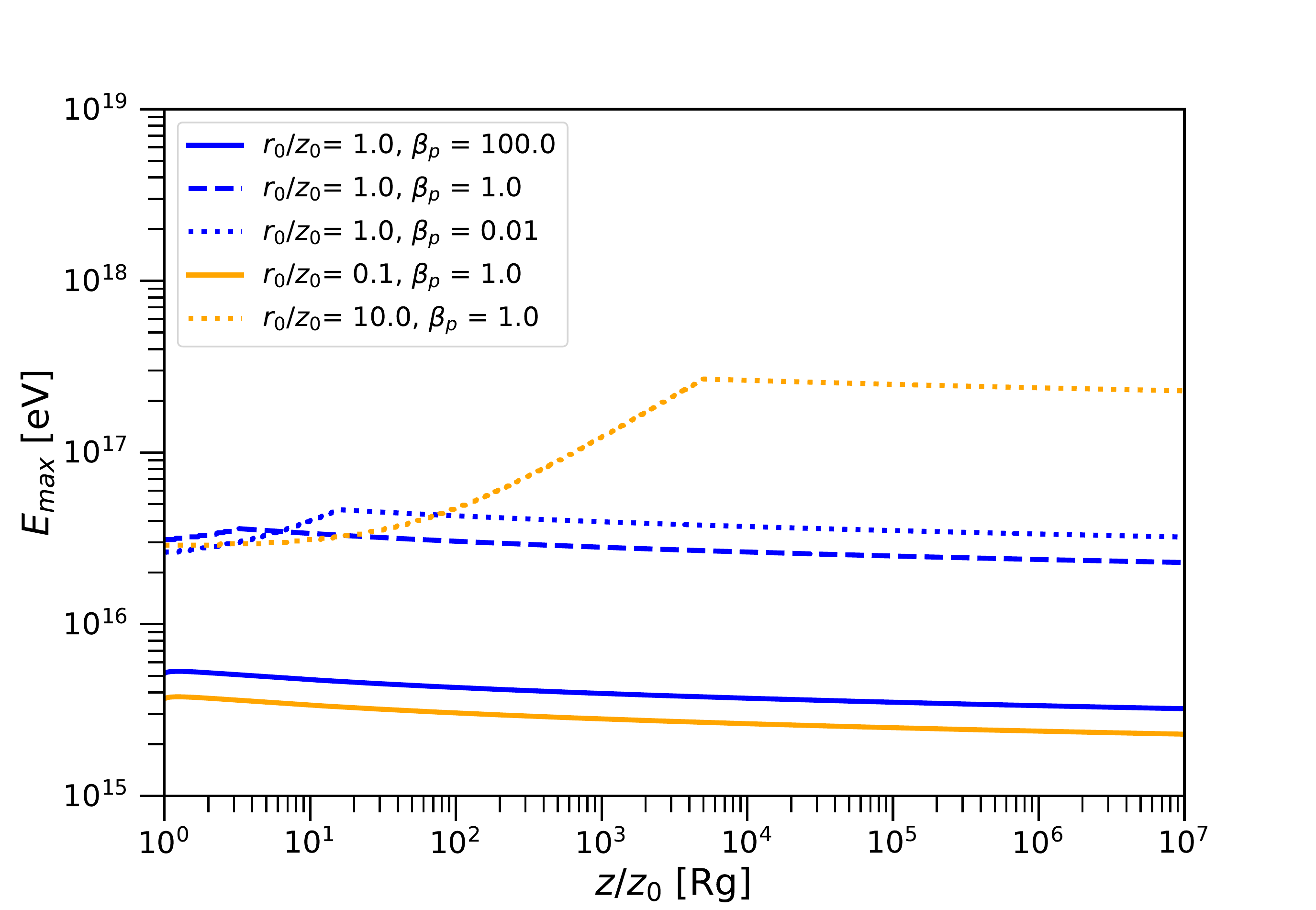}}
\caption{Maximum CR energy as a function of jet height $z$ for the isothermal Jet model.}
\label{fig:IsothermalJetMax}
\end{figure}

\subsection*{Quasi-Isothermal (\textit{agnjet})}
In the quasi-isothermal model, the gas in the jet can only do work in the $z$-direction, meaning that $T_j \propto (\gamma_{j} \beta_j)^{1-\Gamma_{\rm adi}}$. A key difference here is that we include self-collimation, and so the radius of the jet as a function of jet height $z$ is given by Equation \ref{eq:11}. The internal energy density profile is the similar to the isothermal case:
\begin{equation}
    U_j(z) = \zeta n_0 m_p c^2 \bigg(\dfrac{\gamma_j \beta_j}{\gamma_0 \beta_0}\bigg)^{-\Gamma_{\rm adi}} \bigg( \dfrac{r_{\rm coll}}{r_0}\bigg)^{-2}
\end{equation}
Here, we use the collimated radius from Equation \ref{eq:11}. The Euler equation, however includes an additional factor:
\begin{equation}
    \Bigg( \gamma_j \beta_j \dfrac{\Gamma_{\rm adi} + \xi }{\Gamma_{\rm adi} - 1} - \Gamma_{\rm adi} \gamma_j \beta_j - \dfrac{\Gamma_{\rm adi}}{\gamma_j \beta_j} \Bigg) \dfrac{\partial \gamma_j \beta_j }{\partial z} = \dfrac{2}{z}
\end{equation}
\begin{equation}
    \xi = \dfrac{1}{\zeta} \bigg(\dfrac{\gamma_j \beta_j}{\gamma_0 \beta_0}\bigg)^{\Gamma_{\rm adi} -1} 
\end{equation}
The figure for this jet model is found in Section 4, Fig. \ref{fig:QuasiIsothermalMax}. We note that $U_j(z)$ is very similar for the isothermal and quasi-isothermal models, as only the the dependence on the radius is different. For this reason, their maximum CR energies are very similar.





\bibliographystyle{mnras}
\bibliography{references} 





\bsp	
\label{lastpage}
\end{document}